\documentclass[pdflatex,sn-mathphys-num]{sn-jnl}

\usepackage{graphicx}%
\usepackage{multirow}%
\usepackage{amsmath,amssymb,amsfonts}%
\usepackage{amsthm}%
\usepackage{mathrsfs}%
\usepackage[title]{appendix}%
\usepackage{xcolor}%
\usepackage{textcomp}%
\usepackage{manyfoot}%
\usepackage{booktabs}%
\usepackage{algorithm}%
\usepackage{algorithmicx}%
\usepackage{algpseudocode}%
\usepackage{listings}%
\usepackage{comment}

\theoremstyle{thmstyleone}%
\theoremstyle{thmstyletwo}%

\theoremstyle{thmstylethree}%

\begin{document}

\title[Article Title]{Breakloose suppression in minimal friction models}


\author*[1]{\fnm{Shubham} \sur{Agarwal}}\email{shubham.agarwal@uni-saarland.de}



\affil*[1]{\orgdiv{Material Science and Engineering}, \orgname{Universit\"at des Saarlandes}, \orgaddress{\street{Campus, Geb. C6.3}, \city{Saarbr\"ucken}, \postcode{66123}, \state{Saarland}, \country{Germany}}}


\abstract{Breakloose friction—the transient force peak at the onset of sliding—is often pronounced in nanoscale contacts but weak or absent in macroscopic systems. Although this behavior is commonly associated with rupture fronts and process-zone effects, how the stiction peak is controlled by system size, temperature, driving rate, and loading geometry—and what mechanisms underlie its emergence or suppression—remains incompletely understood. Here we investigate this problem using three minimal friction models with distinct loading geometries: a multi-particle Prandtl–Tomlinson system with independently driven particles, an end-driven Frenkel–Kontorova chain with elastic stress transmission along the interface, and a uniformly driven FK chain in which each site is coupled locally to the driving stage. We show that similar macroscopic suppression of breakloose friction can arise from fundamentally different mechanisms. In multi-particle PT systems, increasing system size or temperature promotes statistical dephasing of local depinning events, smoothing the global response. In end-driven FK chains, internal elasticity redistributes stress along the interface, delaying sliding onset and, together with higher temperature or slower driving, enabling progressive relaxation during loading. In uniformly-driven FK chains, the stiffness of the driving springs controls the synchronization of slip events and thereby the character of the sliding response. These results demonstrate that the presence or absence of a breakloose peak does not uniquely identify a single physical mechanism, but instead reflects the interplay of local pinning, elastic coupling, and contact architecture.}




\maketitle

\section{Introduction}\label{sec1}

The force peak observed at the onset of sliding, often referred to as stiction or breakloose friction, is a central feature of frictional systems ranging from nanoscale contacts to macroscopic interfaces. In classical tribology, macroscopic friction originates from the formation and shear failure of microscopic junctions that constitute the real contact area \cite{Bowden2001book}. In many materials these junctions strengthen during stationary contact, so the breakloose force increases with hold time, a phenomenon commonly described using rate-and-state friction laws \cite{Dieterich1978PAG,Dieterich1979JGR, Ruina1983JGR,Marone1998AREPS}.

Beyond contact ageing, the magnitude of the stiction peak also depends strongly on the size and elasticity of the sliding system. Experiments frequently show pronounced stiction in nanoscale contacts, while macroscopic systems often exhibit only weak or no overshoot at the onset of motion. A common explanation invokes a characteristic process-zone length over which the interface transitions from static to kinetic friction during incipient sliding. When the system size greatly exceeds this length, slip can nucleate locally and propagate along the interface through precursor events or rupture fronts, reducing the macroscopic breakloose peak \citep{Palmer1973PRS,Rice1983PAG,Rubinstein2007PRL,Ben2010Science}. Complementary interpretations based on Persson’s multiscale contact mechanics emphasize the role of elastic compliance and progressive interfacial unlocking, which can also suppress breakloose friction under distributed loading conditions \citep{Persson2001JCP,Persson2023JCP}.

Despite these advances, the mechanical origin of stiction suppression remains incompletely understood. In particular, similar macroscopic friction responses can arise in systems with very different microscopic interactions and contact architectures. It is therefore not always clear whether the absence of a pronounced breakloose peak reflects rupture-front dynamics, statistical averaging of local slip events, or progressive stress redistribution within an elastic interface.

\begin{figure}[hbt!]
  \centering
  \includegraphics[width=0.5\textwidth]{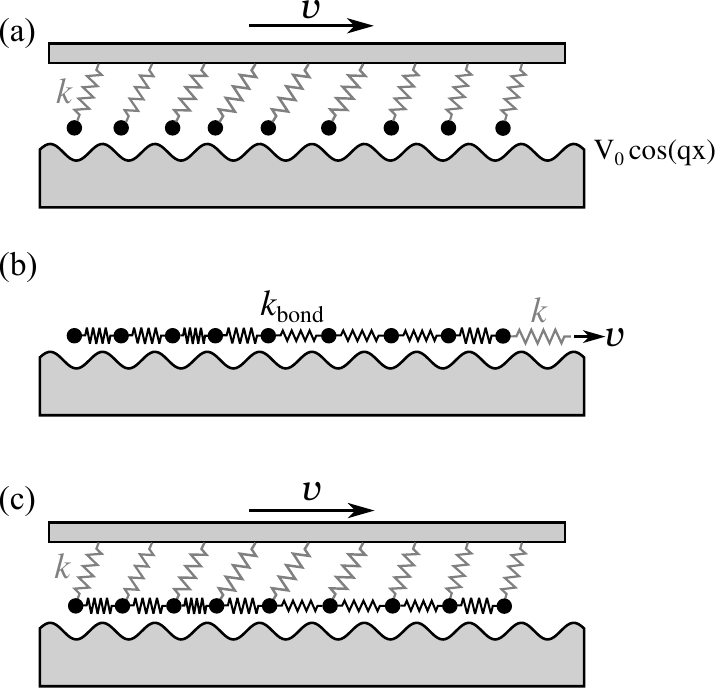}
  \caption{Schematics showing (a) a multi-particle Prandtl-Tomlinson model; (b) an end-driven FK chain; (c) a uniformly-driven FK chain. 
  }
  \label{fig:pt_fk_schem}
\end{figure}

In this paper, we systematically investigate how these distinct mechanisms emerge and interact using minimal models that isolate the roles of elastic coupling, system size, and the manner in which shear is applied. 
We use minimal friction models based on the Prandtl--Tomlinson \cite{Prandtl1928geZAMM,Tomlinson1929ph,Schwarz2016ACS} and Frenkel--Kontorova \cite{Frenkel1938,Braun19981PR} frameworks, which have long served as useful tools for isolating generic mechanisms of frictional sliding \cite{Vanossi2007JP,Norell2016PRE,Pastewka2026TI,Vanossi2008TI}.
Hybrid models like Frenkel--Kontorova--Tomlinson (FKT) have also been studied previously in the context of nanoscale friction and lubricated sliding \cite{Weiss1996PRB, Weiss1997PRB,kim2009PRB}
We employ three canonical friction models representing distinct limiting cases of interfacial loading, as summarized schematically in Fig.~\ref{fig:pt_fk_schem}. 
The first is a multi-particle Prandtl–Tomlinson (MPPT) model, in which the particles are uncoupled and driven independently in the present construction. Such a description is relevant for systems where contact points interact only weakly, for example in atomic force microscopy with multiple independent asperities or in sparse multi-contact interfaces.
The second is an end-driven Frenkel–Kontorova (EDFK) chain, where stress is applied at one boundary and transmitted along the interface through internal elasticity. Similar loading conditions arise in systems where motion is initiated from an edge, such as the peeling of adhesive tapes, the pushing of an object from one side, or rupture propagation along frictional interfaces.
The third is a uniformly driven Frenkel–Kontorova (UDFK) chain, where each site is coupled to the driving stage through a local spring, representing distributed loading conditions. This scenario approximates situations where the interface is driven more uniformly, such as extended contacts between rough surfaces or boundary-lubricated layers sheared between two moving solids.
By comparing these models, we aim to identify the distinct mechanisms responsible for the emergence and suppression of the breakloose peak.


\section{Model and methods}\label{methods}

\subsection{Prandtl-Tomlinson model}
To incorporate finite-size effects, we consider one-dimensional, multi-particle Prandtl-Tomlinson (MPPT) model in which each degree-of-freedom is pulled by a translating stage at velocity $v$ through a spring of stiffness $k$, while moving in a sinusoidal potential of amplitude $V_0$ and wave number $q$. The equation of motion of $n$th particle reads:
\begin{equation}
    m\ddot{x}_n + \frac{m}{\tau}\dot{x}_n + k(x_n-vt)= qV_0\sin{(qx_n + \phi_n)} + \Gamma_n(t)
\end{equation}
where $m$ is the mass of each particle, $\tau$ the relaxation time, and $\phi_n$ the random phase offset of the $n$th contact. The term $\Gamma_n(t)$ represents a thermal random force accounting for finite temperature. It has a zero mean and second moments of $\left\langle \Gamma_m(t) \Gamma_n(t') \right\rangle =  2 k_{\text{B}} T m
\delta_{mn} \delta(t-t') / \tau$, where $k_{\text{B}}T$ is the thermal energy. 
In the present implementation of the MPPT model, three parameters are set to unity, namely bead inertia $m$, the corrugation barrier $V_0$, and the lattice wave number $q=2\pi/a$, to define our unit system.
For the default model in this study, we keep $k/(qV_0)=0.1$, $k_{\text{B}}T=0.1V_0$, and driving speed of $v=0.001$.
In the present construction, the contacts are uncoupled, so the macroscopic response arises from the statistical superposition of many independent depinning events. This makes the MPPT model a useful limiting case for weakly interacting contacts, while omitting spatially extended effects such as edge-mediated loading, finite-contact stress redistribution, and collective interfacial rearrangements that may arise in extended contacts~\cite{deWijn2012PRB}.

\subsection{Frenkel-Kontorova model}
To introduce internal elastic degrees of freedom along the interface, we consider a one-dimensional Frenkel-Kontorova model \cite{Frenkel1938,Braun19981PR} consisting of a linearly elastic bead-spring chain with nearest-neighbor interactions, which is placed into a single-sinusoidal potential of the form $V(x) = V_0 \cos(qx)$. The equation of motion for $n$th particle in a $N$-particle FK chain can be written as: 
\begin{equation}
\label{eq:fk_model}
m \ddot{x}_n + \frac{m}{\tau} \dot x_n + 2k_{\text{bond}} \left(x_n-\frac{x_{n+1} + x_{n-1}}{2}\right) =
q V_0 \sin(qx_n)+ F_n^{\mathrm{drive}}(t)+ \Gamma_n(t).
\end{equation}
where $m$ is the mass of each particle, $\tau$ the relaxation time, and particle index $n= 0,1,...,N-1$. The driving force is defined as:
\begin{equation}
F_n^{\mathrm{drive}}(t) =
\begin{cases}
k(vt+x_{N-1}(0) - x_{N-1})\,\delta_{n,N-1}, & \text{EDFK},\\[0.4em]
k(vt+x_n(0) - x_n), & \text{UDFK},
\end{cases}
\end{equation}
The two terminating beads are treated depending on the boundary condition.
End beads of an open chain are only coupled to one neighbor.
In the case of periodic boundary conditions (PBC), the first and last bead subject a force of $F_{0,N-1} = \mp k (x_{0} - x_{N-1} + L - b)$ onto each other, where $b = L/N$ would be the preferred spacing beads in a free chain, just like in an open chain. 
In the current implementation of the FK, the three parameters set to unity are: bead inertia $m$, the corrugation barrier $V_0$, and the equilibrium bead-spacing $b$.
For the default FK model in this study, we consider a chain of 14 atoms interacting with 15 substrate wells, and set $k_{\text{bond}}=10V_0/b^2$, $k=0.1k_{\text{bond}}$. The chosen $k_{\text{bond}}$ corresponds to a moderately coupled chain \citep{Agarwal2025PRE}.

The above two equations of motion are solved using an in-house written MD Python code. We choose $\tau=(2\pi/q)\sqrt{m/V_0}$ and use the thermostat-scheme by the Grønbech-Jensen~\cite{GronbechJensen2019MP} for our simulations.
\section{Results}\label{results}

\subsection{Prandtl-Tomlinson model}

\begin{figure}[hbt!]
  \centering
  \includegraphics[width=0.475\textwidth]{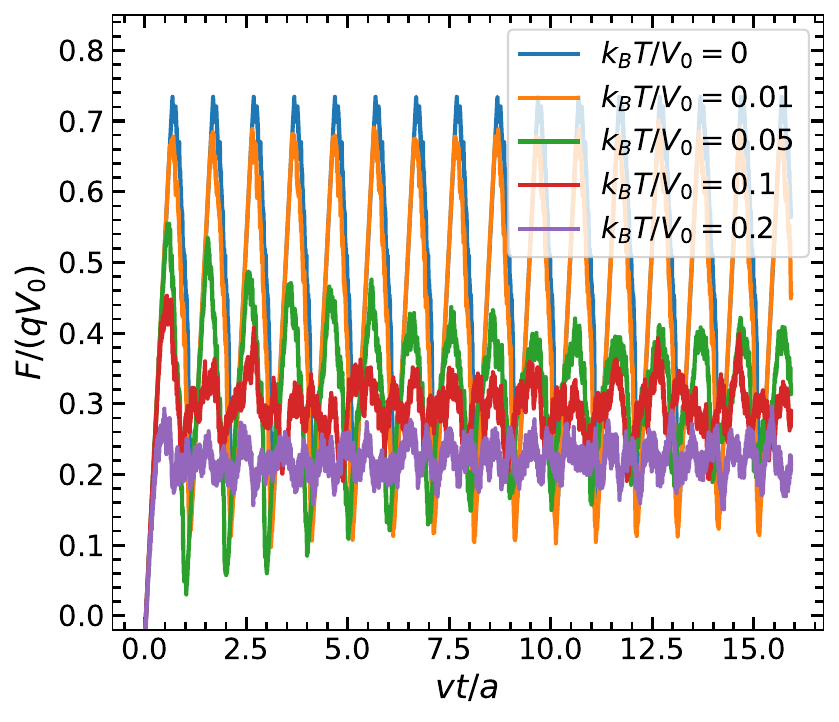}\hfill
  \includegraphics[width=0.495\textwidth]{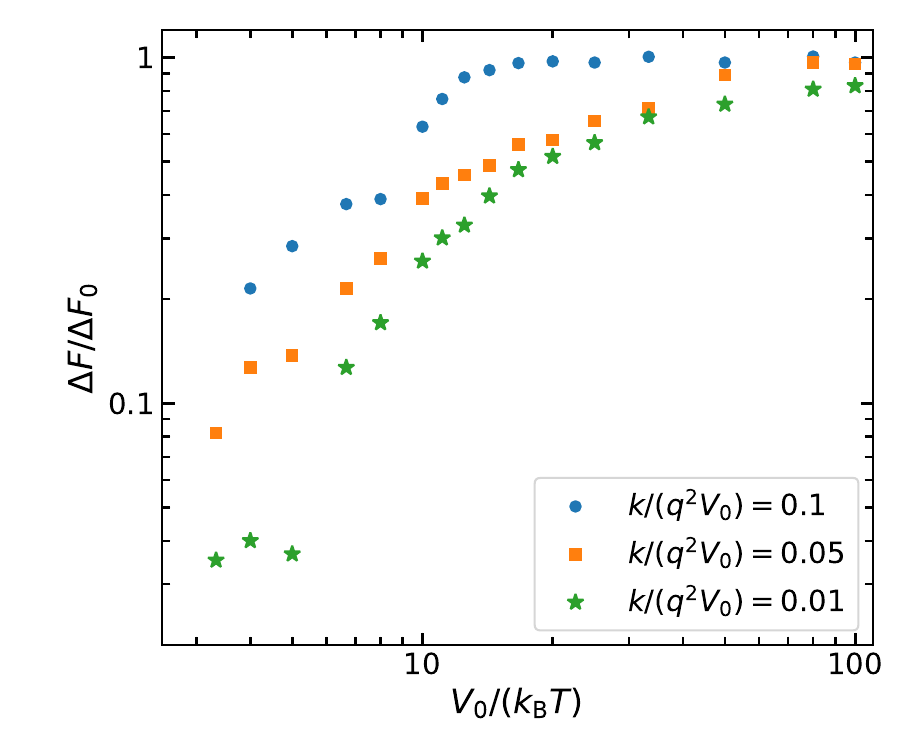}
  \caption{The effect of temperature for a MPPT model with 128 atoms on (Left) force trace, and (Right) stiction peak magnitude. 
  }
  \label{fig:pt_therm}
\end{figure}

Fig.~\ref{fig:pt_therm} shows the effect of temperature in the multi-particle Prandtl–Tomlinson (MPPT) model \cite{Prandtl1928geZAMM,Tomlinson1929ph,Popov2014bookch}. The left panel plots the sliding force as a function of sliding distance. In the athermal (low $T$) limit, the force signal displays pronounced, highly regular stick–slip, reflecting abrupt and nearly synchronous depinning from successive metastable substrate wells. With increasing temperature, thermally activated barrier crossing triggers earlier slip of individual particles, progressively desynchronizing depinning events and spreading force drops in time; consequently, the macroscopic force trace smooths and the sharp oscillations give way to a broader breakloose maximum at sliding onset, followed by irregular—or at sufficiently high $T$, nearly smooth—sliding. The right panel quantifies this trend for three spring stiffness $k$ via the stiction peak amplitude 
$\Delta F=F_p-F_k$,
where $F_p$ denotes the peak force at sliding onset and $F_k$ the subsequent steady sliding (kinetic) friction level. The athermal stiction peak amplitude is denoted by $\Delta F_0$.
As temperature decreases, $\Delta F$ increases and eventually saturates in the low-$T$ limit, consistent with a crossover from thermally activated depinning to the deterministic instability-controlled regime. Varying the driving stiffness shifts the magnitude of the response without changing these qualitative trends: stiffer springs store more elastic energy prior to depinning and therefore produce a larger $\Delta F$ across the entire temperature range.

\begin{figure}[hbt!]
  \centering
  \includegraphics[width=0.5\textwidth]{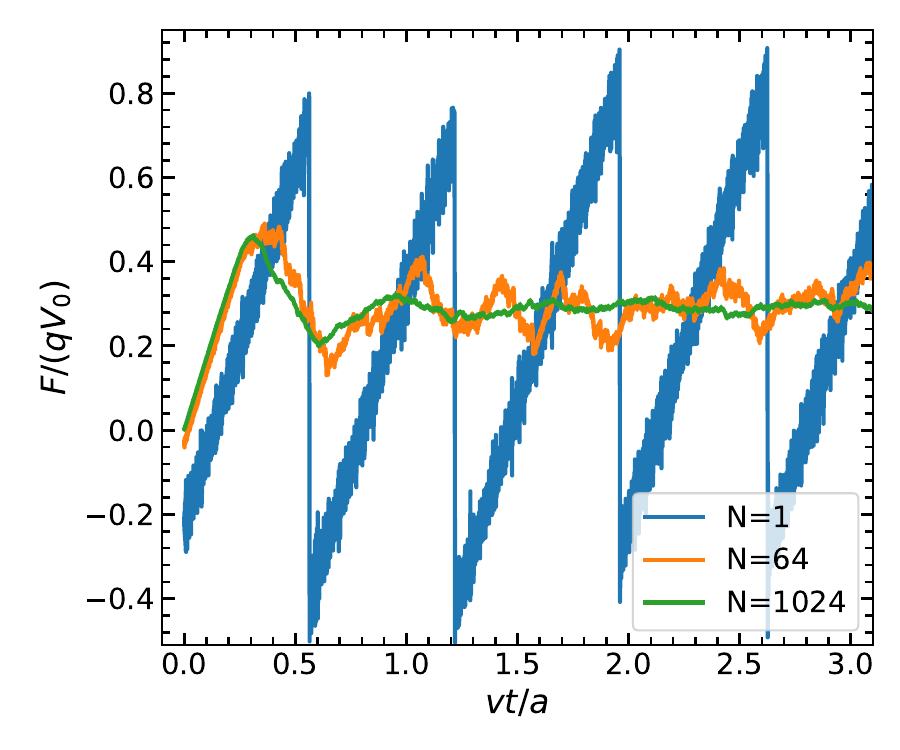}\hfill
  \includegraphics[width=0.5\textwidth]{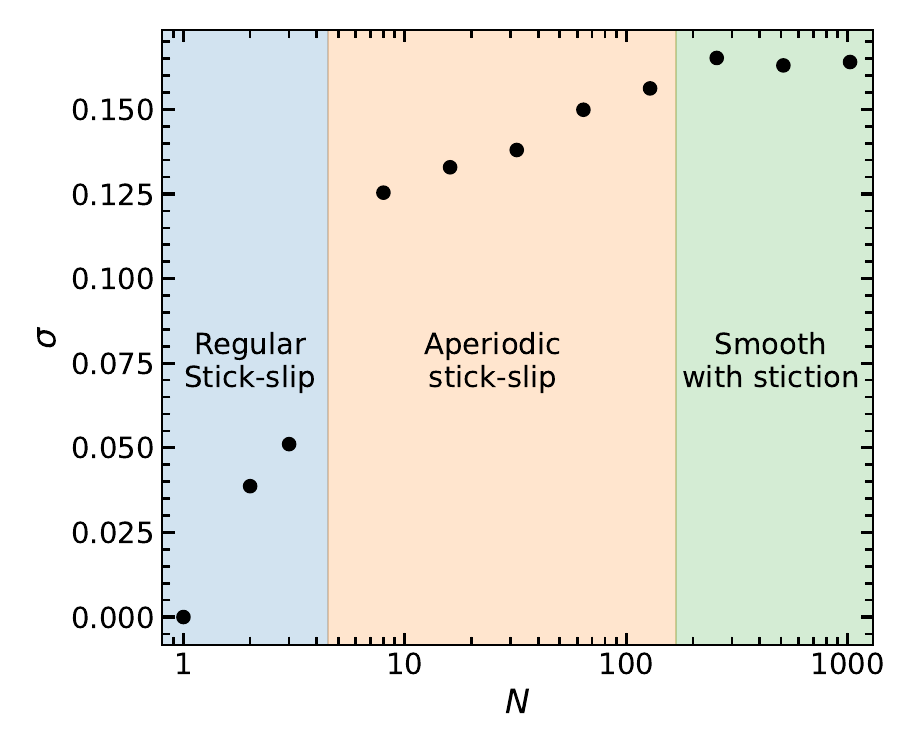}
  \caption{ (Left) The effect of system-size on force trace per particle for a 1D-PT model at $k_BT/V_0=0.1$; (Right) The standard deviation in force-per-particle at macroscopic stiction peak as a function of system size. 
  }
  \label{fig:pt_size}
\end{figure}

Fig.~\ref{fig:pt_size} shows the effect of system size at fixed temperature. The left panel presents the total force signal for different number of particles $N$, while the right panel quantifies the particle-to-particle force dispersion at sliding onset, $\sigma$, which serves as a measure of stick–slip coherence. A clear trend emerges: as $N$ increases, individual depinning events become increasingly desynchronized, so force drops are distributed over time and the global force trace progressively smooths, suppressing coherent stick–slip. The response crosses over through three regimes: for small systems ($N < 4$), depinning remains highly synchronized, producing regular stick–slip and a rapid increase of $\sigma$ with $N$; for intermediate sizes ($4 < N < 150$), partial dephasing leads to an aperiodic stick–slip and a more gradual $\sigma(N)$ growth; for large systems ($N>150$), the macroscopic signal is largely smooth (with a stiction overshoot) and $\sigma$ saturates, consistent with fully dephased, effectively uncorrelated slip. 
These results show that even without internal elasticity, increasing system size suppresses macroscopic instabilities via statistical dephasing of local depinning events. A similar elimination of stick–slip by partitioning a rough/split interface into many sub-contacts was reported by \citep{Kligerman2014TL}.

\subsection{End-driven Frenkel-Kontorova chain}
We now consider a one-dimensional end-driven Frenkel--Kontorova (EDFK) chain, in which the interface is pulled from one end through a spring attached to a translating stage. In contrast to the multi-contact PT model, the FK chain includes elastic coupling between neighboring degrees of freedom and therefore permits stress redistribution along the interface.

\begin{figure}[hbt!]
  \centering
  \includegraphics[width=0.5\textwidth]{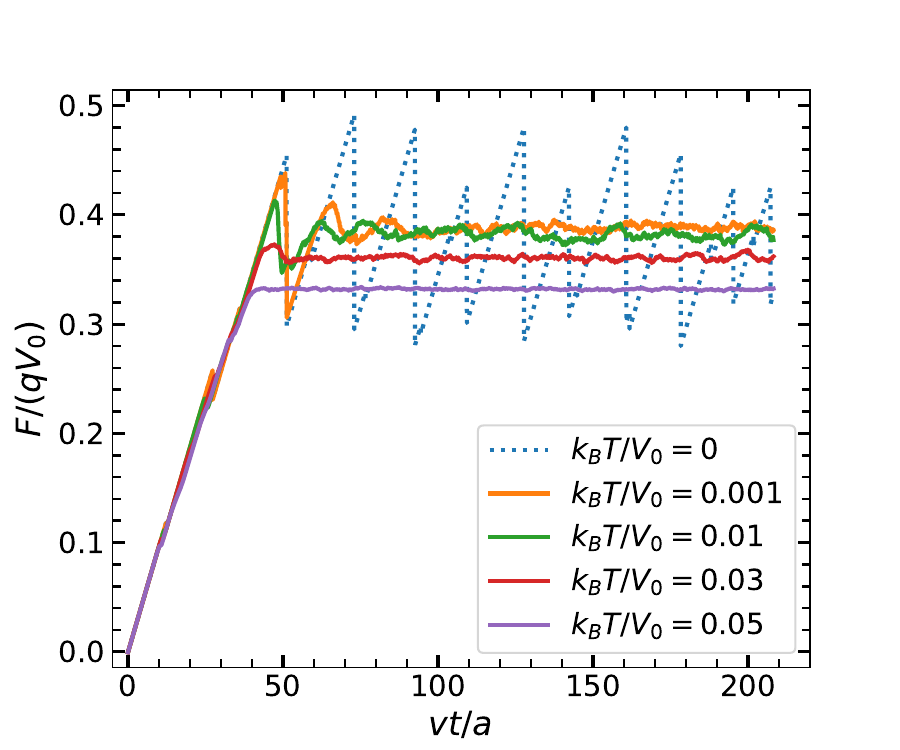}\hfill
  \includegraphics[width=0.5\textwidth]{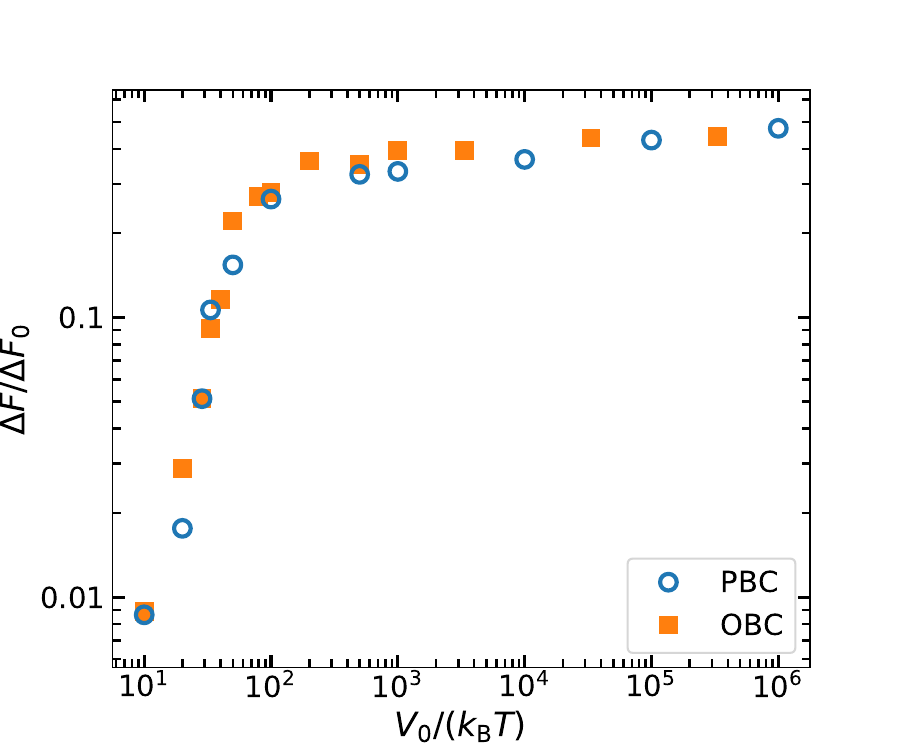}
  \caption{(Left) The effect of temperature on the evolution of friction-force per particle for periodic 1D-FK chain; (Right) The variation of stiction peak magnitude ($\Delta F$) with temperature for periodic and open chains.  Chain is end-driven with soft spring ($k/k_{\text{bond}}=0.1$) at $v=1\times 10^{-3}$.}
  \label{fig:fk_therm_bc}
\end{figure}

Fig.~\ref{fig:fk_therm_bc} shows the effect of temperature in the end-driven FK chain. A clear  crossover is observed in the character of the onset of sliding. At low temperatures ($k_{\text{B}}T\leq0.03$), the force–displacement response exhibits a distinct stiction (breakloose) peak, whereas at higher temperatures the peak progressively weakens and ultimately disappears. Increasing temperature promotes barrier crossing and local rearrangements ahead of macroscopic motion, allowing the interface to relax shear stresses gradually so that the onset becomes smooth and the force approaches the kinetic level without an overshoot. 

\begin{figure}[hbt!]
  \centering
  \includegraphics[width=0.5\textwidth]{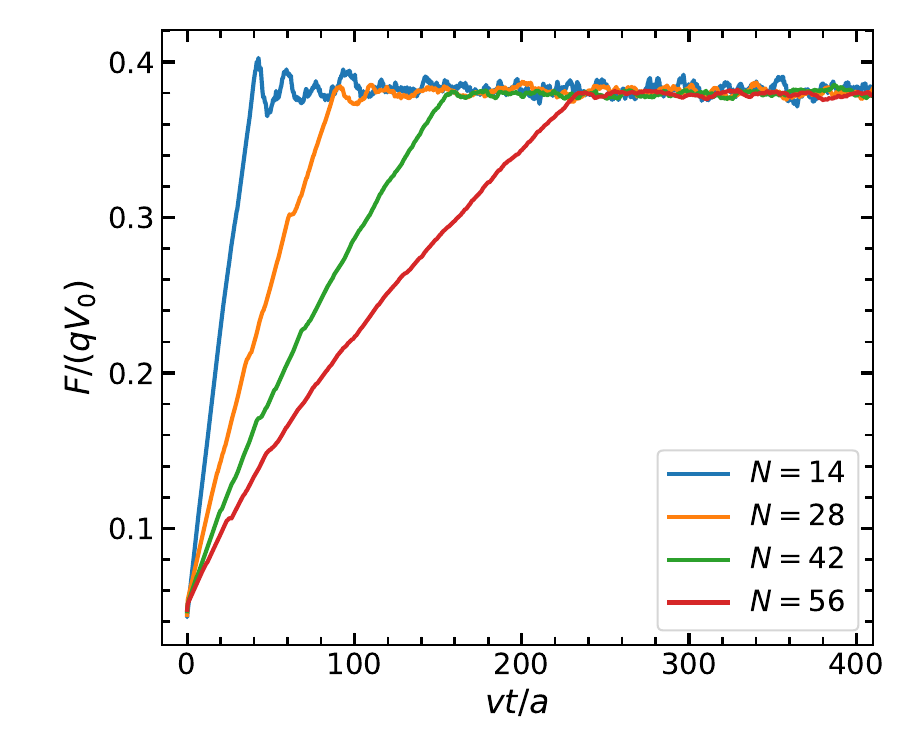}\hfill
  \includegraphics[width=0.5\textwidth]{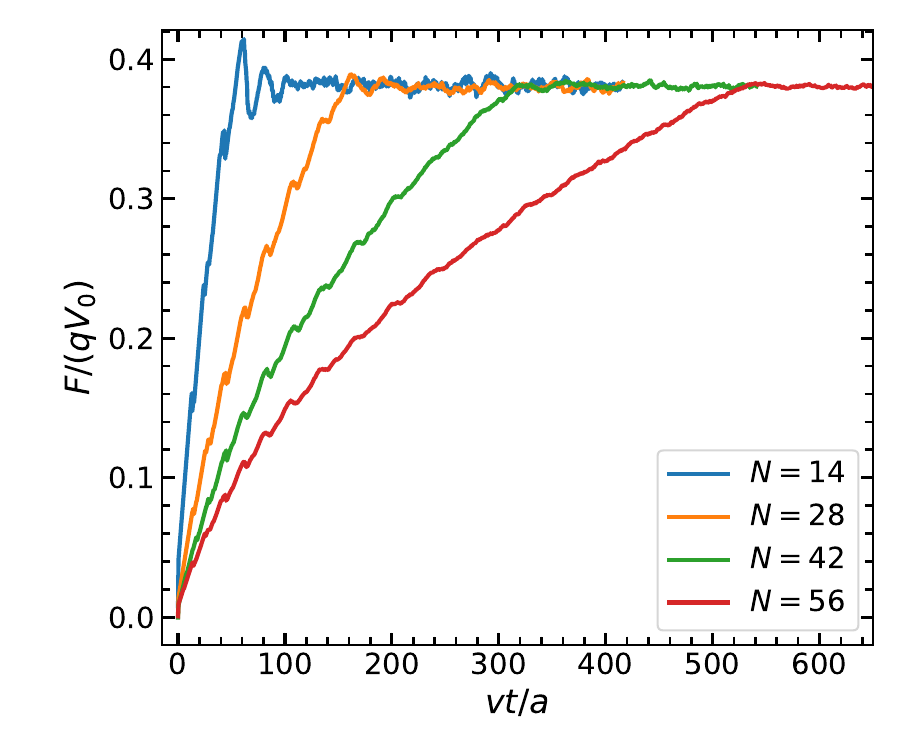}
  \caption{The effect of chain length on friction-force per particle for an end-driven 1D-FK chain with periodic (left) and open (right) boundary conditions. End particle of FK-chain is pulled by a soft spring ($k_{\text{drive}}/k_{\text{spring}}=0.1$) at $v=1\times 10^{-2}$.}
  \label{fig:fk_size_bc}
\end{figure}

Fig.~\ref{fig:fk_size_bc} shows how chain length influences breakloose behavior in EDFK chains with both periodic and open boundary conditions. 
While the steady-state sliding force per particle converges to a size-independent plateau, the transient response exhibits a pronounced system-size dependence. Short chains (e.g., $N=14$) show a distinct stiction peak associated with abrupt depinning. As the chain length increases, however, this peak is progressively suppressed and the onset of sliding is delayed. 
Small precursor slips can be observed in the force response prior to the major breakloose event. Such precursor activity has also been observed in PMMA sliding experiments \citep{Katano2014SR}. These minor slips are signatures of progressive stress relaxation and redistribution within the contact interface, thereby reducing the magnitude and abruptness of the major depinning event.
This interpretation is also consistent with earlier FK-based descriptions of the transition from static to dynamic friction, in which slip is mediated by rupture-like fronts propagating along the interface \citep{Gershenzon2013TI}.
%
In the present system, this behavior arises naturally from the internal elasticity of the chain and the effect is particularly pronounced for open boundary conditions.
In longer chains, multiple precursor slips are observed. These precursor events become increasingly frequent but smaller in magnitude, indicating progressive relaxation of the accumulated elastic stress. As a result, the transition from static to kinetic sliding becomes progressively smoother with increasing system size.
This size effect is fundamentally distinct from the statistical dephasing observed in MPPT where increasing the number of particles smooths the signal through averaging over independent slip events without affecting the timing of onset. 
In contrast, EDFK chains modify the stress distribution itself, so that the loss of coherence arises from elastic stress redistribution and precursor activity within the chain.
%

%

\begin{figure}[hbt!]
  \centering
  \includegraphics[width=0.5\textwidth]{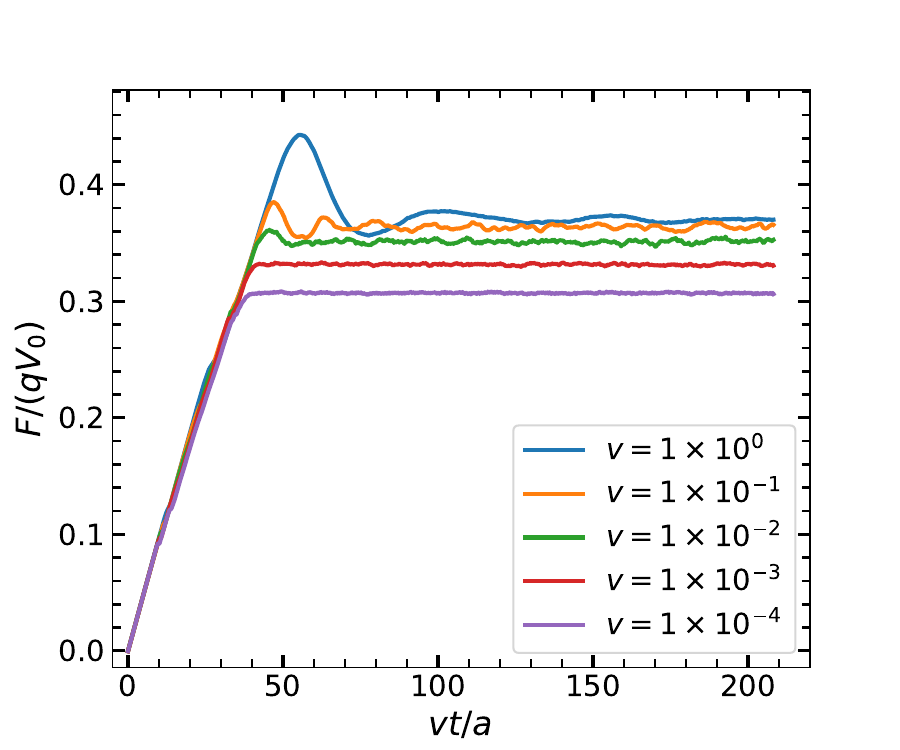}\hfill
  \includegraphics[width=0.5\textwidth]{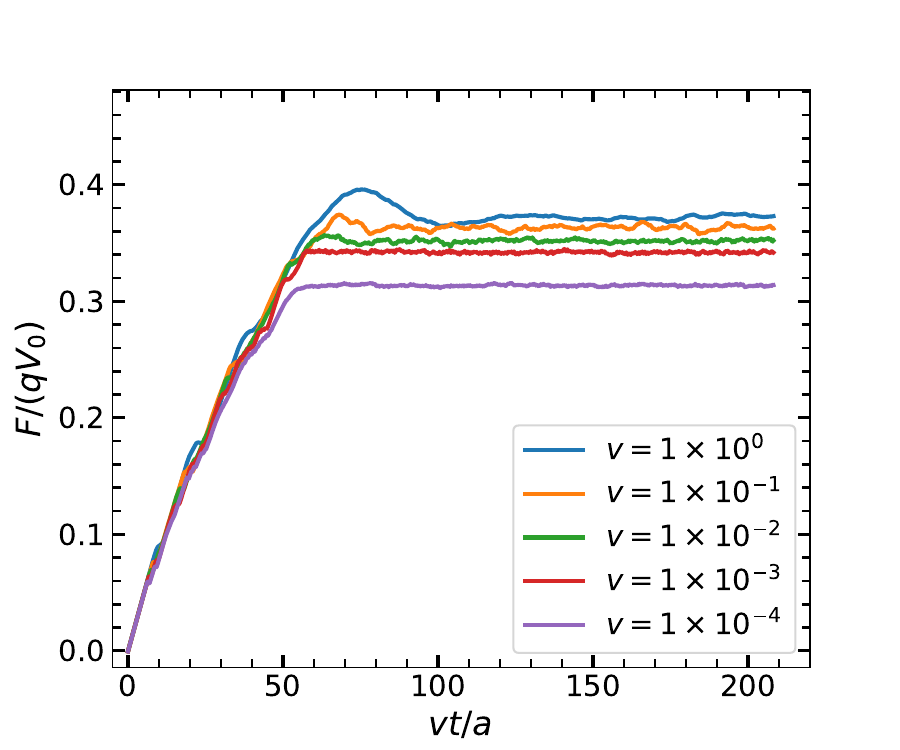}
  \caption{The effect of pulling rate $v$ on friction-force per particle for an end-driven 1D-FK chain with periodic (left) and open (right) boundary conditions. End particle of FK-chain is pulled by a soft spring ($k/k_{\text{bond}}=0.1$) at $k_{\text{B}}T=0.05$.}
  \label{fig:fk_rate_bc}
\end{figure}

We now examine the effect of driving rate on the onset of sliding in an end-driven periodic and open FK chains at finite temperature which are shown in Fig.~\ref{fig:fk_rate_bc}. At higher pulling speeds, the friction force exhibits a pronounced stiction peak, indicating a relatively abrupt depinning transition. As the driving rate is reduced toward the quasi-static regime, this peak weakens and can disappear altogether, and the force signal becomes smoother. This trend is consistent with a competition of timescales: at slow driving the system has sufficient time to undergo thermally assisted local rearrangements during the loading phase, which redistributes stress and releases stored elastic energy progressively, thereby suppressing a coherent breakloose event. At fast driving, these relaxation pathways are outpaced, so elastic energy accumulates and is released more abruptly, producing a measurable overshoot.  Similar signatures of reduced friction and uniform interfacial stress-distribution at slow sliding speeds are also observed in experiments \citep{Lee2013PNAS} and more realistic MD simulations \citep{Sukhomlinov2025RPP}.

\begin{figure}[hbt!]
  \centering
  \includegraphics[width=0.98\textwidth]{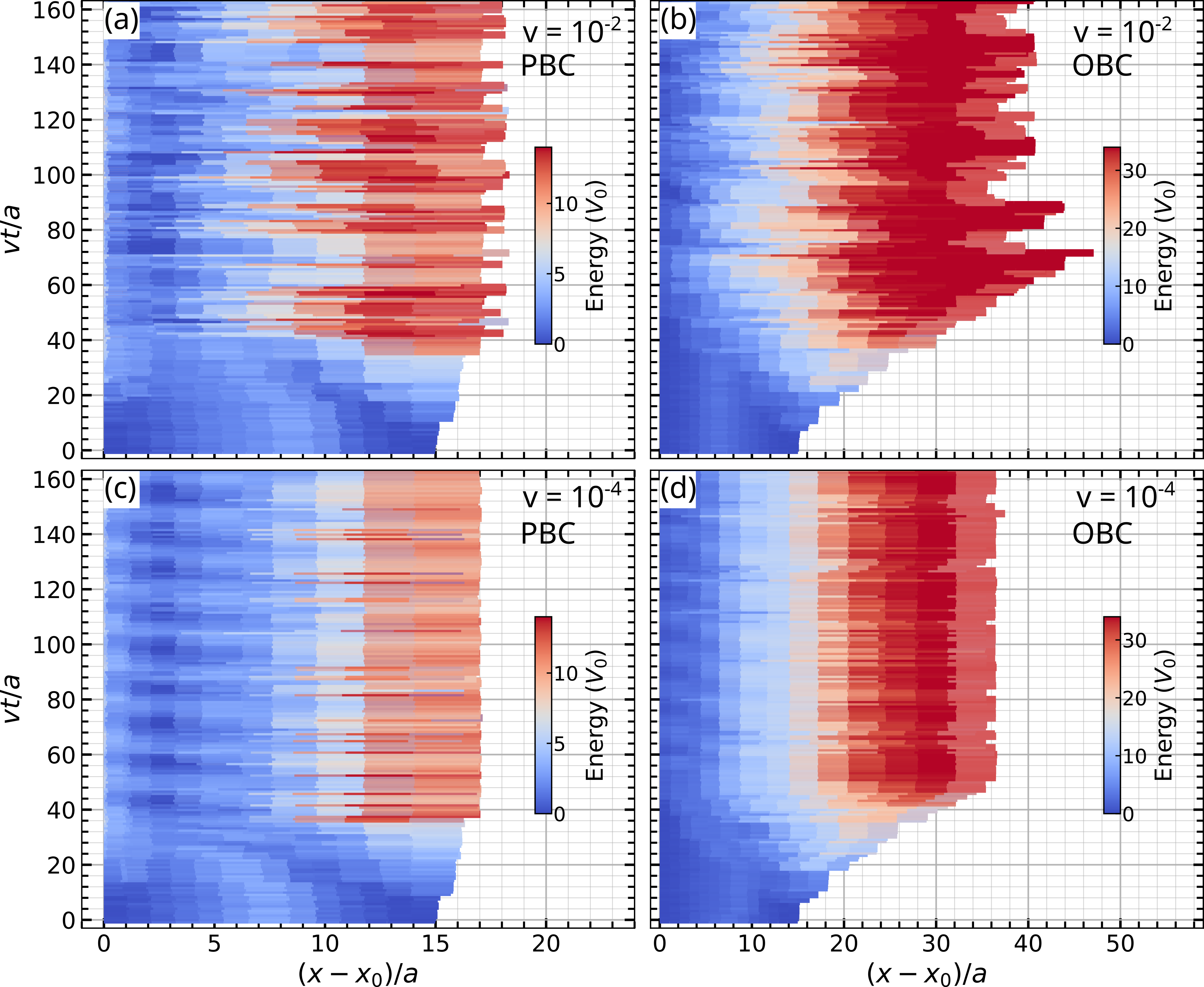}
  \caption{Spatiotemporal maps for: (a) $v=10^{-2}$, PBC; (b) $v=10^{-2}$, OBC; (c) $v=10^{-4}$, PBC; (d) $v=10^{-4}$, OBC. Chains are driven with a soft spring ($k_{\text{drive}}/k_{\text{spring}}=0.1$) at $k_{\text{B}}T=0.05$. The colorbar shows the total chain energy per particle. The horizontal coordinate is plotted as ($(x-x_0)/a$), where $x_0$ is the position of first particle, so that chain is shown comoving with the left end. Fast driving produces a more abrupt depinning event, whereas slow driving leads to more gradual, spatially distributed loading. Open boundaries allow larger pre-slip stretching before sliding onset than periodic boundaries.}
 \label{fig:fk_rate_config}
\end{figure}

Finally, the configuration visualizations in Fig.~\ref{fig:fk_rate_config} substantiate the above interpretations by showing how stress and deformation evolve in space and time. 
Comparing fast and slow driving, the fast driving panels (top) show a more abrupt transition consistent with a sharper depinning event, whereas slow driving (bottom) reveal a more gradual, spatially distributed evolution consistent with progressive accommodation during loading.
The visualizations further clarify the role of boundary conditions: for the same driving rate and temperature, the open chain (right) develops a larger pre-slip stretch before sliding-onset than the periodic chain (left), indicating that open boundaries provide additional compliance and relaxation pathways that prolong the loading transient. Periodic closure reduces these pathways and accelerates the onset, consistent with the shorter delay observed in the force traces under PBC in Fig.~\ref{fig:fk_size_bc}(left).

\subsection{Uniformly-driven Frenkel-Kontorova chain}
We now analyze the uniformly-driven FK model (UDFK) and identify the mechanisms responsible for its characteristic response, emphasizing both the commonalities and the deviations from the previously discussed loading geometries.
In the uniformly driven FK model, each particle is connected to the translating stage through its own spring of stiffness $k$. The external loading is therefore uniformly distributed over the entire chain, with the total driving stiffness scaling as $Nk$. This construction differs from the end-driven case in that shear is applied locally at every site rather than transmitted from a single boundary.
The UDFK model is closely related to hybrid FK–Tomlinson–type models. It also resembles the Burridge–Knopoff model \cite{Burridge1967BSSA}, except that the phenomenological dry-friction law is replaced here by an explicit on-site interaction with the lower body.

\begin{figure}[hbt!]
  \centering
  \includegraphics[width=0.5\textwidth]{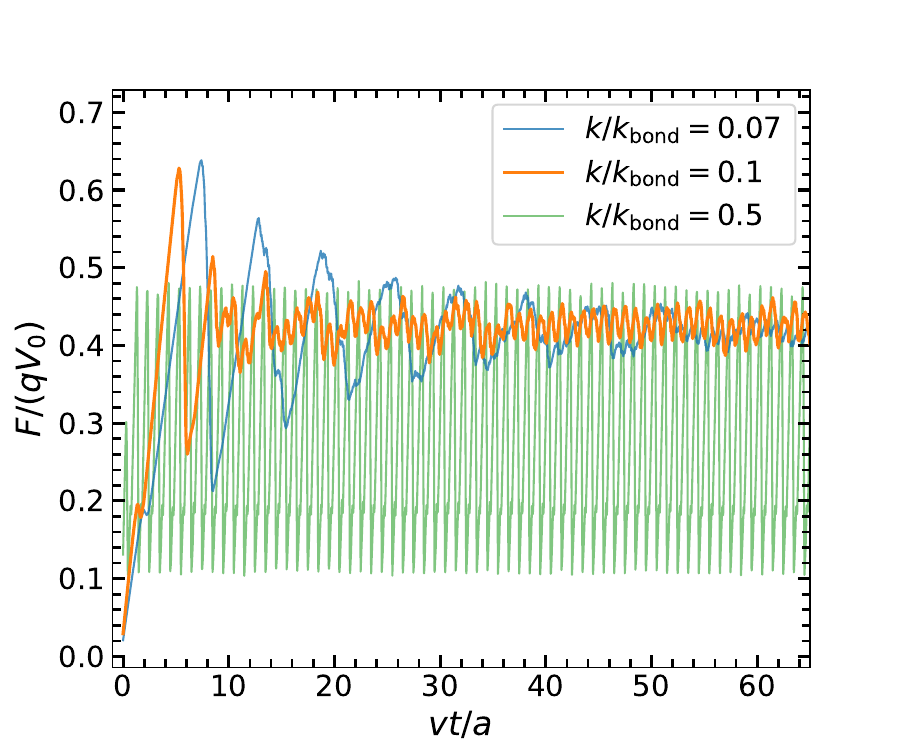}\hfill
  \includegraphics[width=0.5\textwidth]{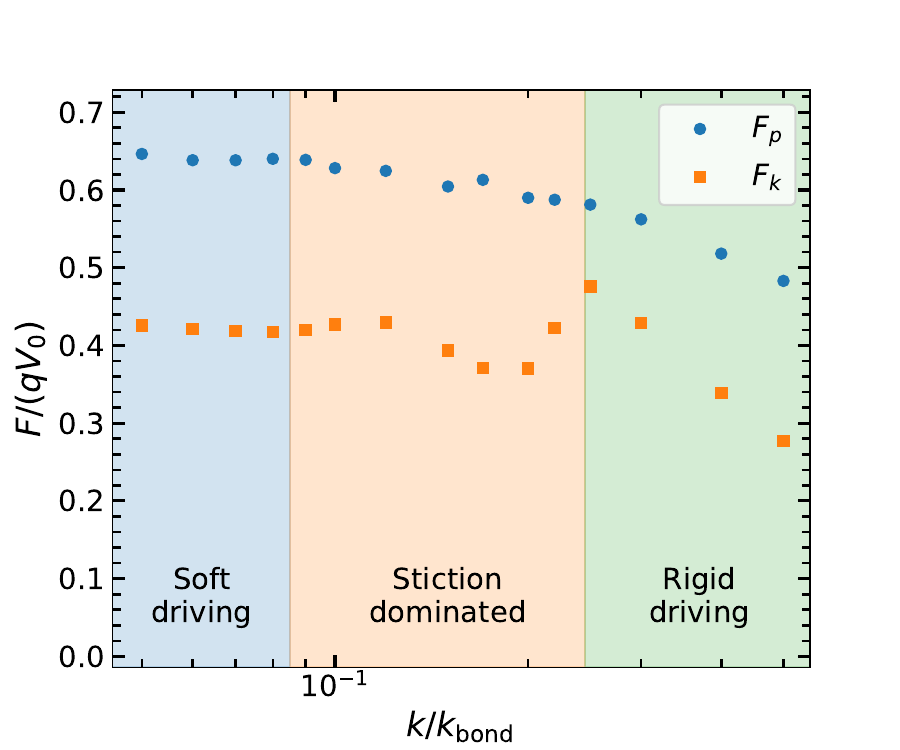}
  \caption{The effect of driving stiffness $k$ on (left) evolution friction-force per particle and (right) peak force ($F_p$) and kinetic friction force ($F_k$) for a uniformly-driven 1D-FK chain with open boundary conditions. The chain is driven at $v=10^{-3}$ at $k_{\text{B}}T=0.05$.}
  \label{fig:fk_multi_obc_k}
\end{figure}

As shown in Fig~\ref{fig:fk_multi_obc_k}, in uniformly driven FK chains the frictional response depends sensitively on the stiffness $k$ of the driving springs. For larger $k$, the loading is effectively rigid and spatially uniform, enabling internal elastic coupling to coordinate slip; the force builds up more steeply, the breakloose force is reduced, and stick–slip persists in steady sliding. 
As $k$ decreases, a pronounced stiction peak emerges and the system relaxes rapidly to its steady sliding level. For even softer driving, the transient response becomes more gradual and the force decays more slowly toward the steady state.
Thus, the driving stiffness acts as a synchronization parameter, controlling whether slip occurs cooperatively across the chain or through localized stiction–relaxation events. 
In contrast, for the end-driven FK chain, varying the pulling stiffness mainly shifts the onset time of slip without qualitatively changing the sliding dynamics.
This highlights that the interplay between internal elasticity and loading stiffness in UDFK can control not only the onset but also the character of frictional sliding, which may be relevant for systems with compliant interfacial layers \citep{Lee2013PNAS,Rahat2025TI} and for models of earthquake rupture dynamics \citep{Tinti2016JGR}.

\begin{figure}[hbt!]
  \centering
  \includegraphics[width=0.99\textwidth]{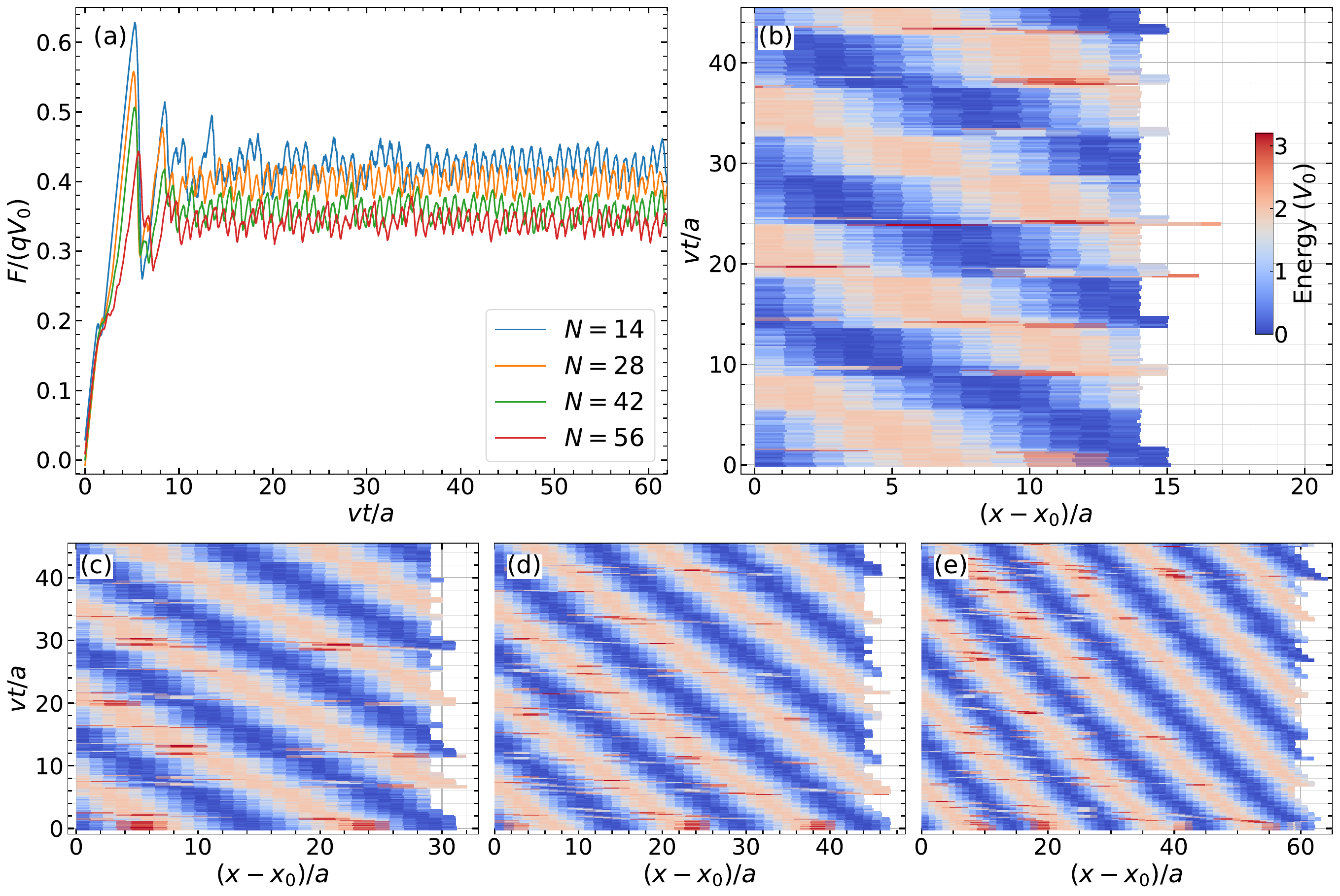}
  \caption{The effect of chain-length $N$ on (a) evolution of friction-force per particle for open UDFK chain; (b--e) Spatiotemporal dissipation maps for $N=14,~28,~42,~\text{and}~56$. The chain is driven at $v=10^{-3}$ at $k_{\text{B}}T=0.05$. In panels (b--e), the horizontal coordinate is plotted as $(x-x_0)/a$, where $x_0$ is the position of first particle, so that chain is shown co-moving with the left end.}
  \label{fig:fk_multi_obc_size}
\end{figure}

Fig~\ref{fig:fk_multi_obc_size} shows that increasing the chain length in the uniformly driven FK model reduces both the breakloose peak and the mean steady friction level, while some stick–slip persists at all sizes. This is in clear contrast to the EDFK chain, whose response becomes much smoother because the dynamics is controlled by a single boundary-driven stress-redistribution process. In the uniformly driven case, the accompanying space-time maps show that increasing 
$N$ does not enlarge one dominant front-like zone, but instead increases the number of distributed stress-gradient regions across the interface. These broader loading bands also contain localized hot spots, indicating short-lived local rearrangements during sliding. The smoother motion of the left side in longer chains further suggests that internal strain is accommodated more gradually between major instabilities. Thus, longer UDFK chains accommodate the imposed shear through many simultaneously loaded and intermittently relaxing domains, so that the macroscopic response reflects the superposition of several local stick–slip cycles rather than one dominant global event. This distributed stress accommodation keeps the interface, on average, less highly loaded, thereby lowering both the mean friction and the macroscopic breakloose peak. However, stick–slip is not eliminated altogether. Rather, the local loading–release cycle remains operative, but becomes partitioned among many coexisting domains across the interface.

\section{Conclusions}\label{conclusion}

In this work, we investigated how the breakloose peak emerges or is suppressed in minimal friction models representing different kinds of contacts and loading conditions. Although all three models can exhibit similar dependence on various parameters, the underlying mechanisms are not the same.

In weakly coupled contacts, represented here by the multi-particle Prandtl– Tomlinson model, suppression of the breakloose peak arises from statistical dephasing: as the number of independently slipping units increases, local depinning events become increasingly desynchronized and the macroscopic force response smooths out.
In boundary-driven elastic contacts, represented by the end-driven Frenkel–Kontorova chain, suppression occurs through stress redistribution along the interface, which delays sliding onset and promotes precursor-like relaxation before global motion.
In uniformly driven distributed contacts, represented by the uniformly driven Frenkel– Kontorova chain, suppression arises through distributed stress accommodation: increasing system size partitions the interface into many simultaneously active loading and relaxation domains, lowering both the mean friction and the macroscopic breakloose peak while preserving local stick–slip dynamics.

The mechanisms identified here should be viewed as complementary to contact-ageing scenarios, in which the evolution of interfacial strength during stationary contact can play an additional role in determining the breakloose response.

\backmatter

\bmhead{Acknowledgements}
The author gratefully acknowledges Prof. Dr. Martin H. M\"user for his insightful suggestions.


\begin{appendices}

\section{MPPT: Driving-rate effect}\label{secA1}

\begin{figure}[hbt!]
  \centering
  \includegraphics[width=0.5\textwidth]{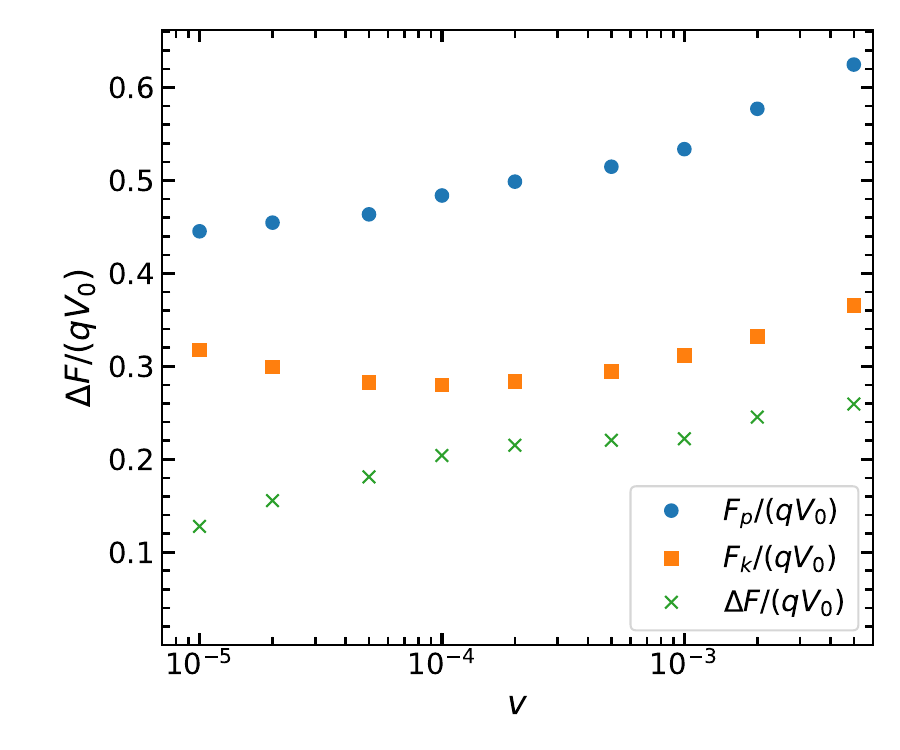}\hfill
  \includegraphics[width=0.5\textwidth]{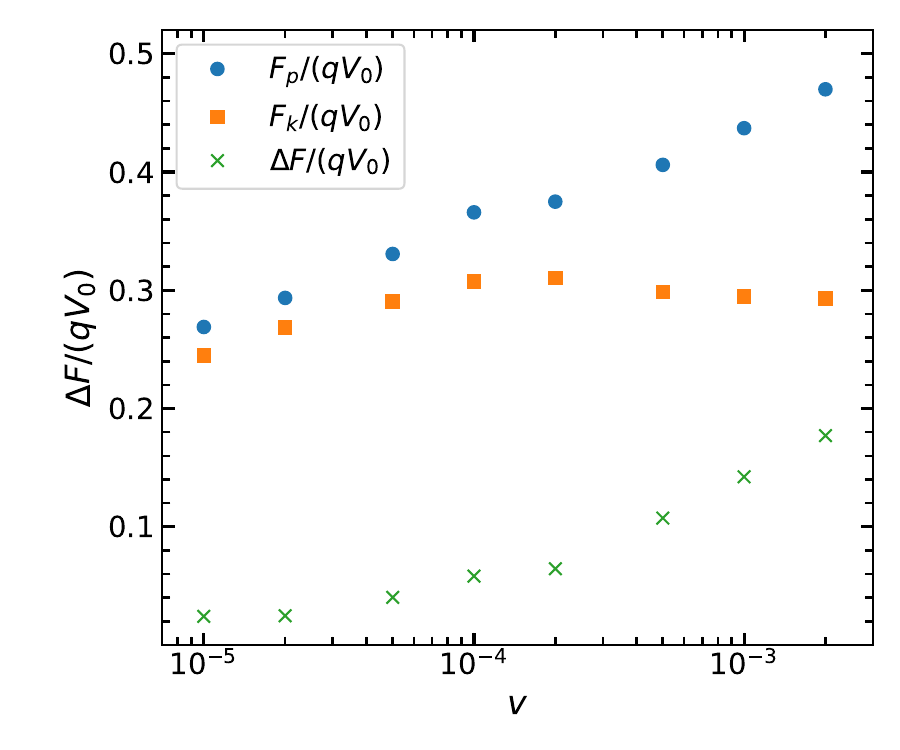}
  \caption{ The effect of driving velocity on peak force, mean steady force, and stiction peak magnitude for (left) $k_{\text{B}}T=0.05V_0$ and (right) $k_{\text{B}}T=0.1V_0$.
  }
  \label{fig:pt_rate}
\end{figure}

Fig~\ref{fig:pt_rate} shows the effect of driving velocity on the peak force $F_p$, the steady sliding force $F_k$, and the stiction magnitude $\Delta F = F_p - F_k$ for two representative temperatures. In both cases, $F_p$ increases with velocity, indicating that faster driving suppresses thermally activated early depinning and shifts slip closer to the deterministic instability. By contrast, $F_k$ exhibits a weaker, non-monotonic, and temperature-dependent rate response, with apparent velocity-strengthening and weakening regimes within the range studied. Consequently, the increase of $\Delta F$ with velocity is governed mainly by the rate dependence of $F_p$, showing that driving rate primarily affects the loading dynamics preceding slip rather than the steady sliding state. A detailed discussion of the velocity dependence of friction in the PT model lies beyond the scope of the present work; we only note that the PT framework exhibits a nontrivial friction–velocity relation across regimes \cite{Muser2011PRB}. Interestingly, related work has also pointed out a formal similarity between the velocity dependence of Prandtl-type friction and rheological shear-thinning in liquids \cite{Gao2024TL}.

\end{appendices}




\end{document}